\documentclass[a4paper, 12pt, UKenglish]{article}

\usepackage[utf8]{inputenc}
\usepackage{gensymb}
\usepackage{amsmath}
\usepackage[parfill]{parskip}           
\usepackage[a4paper,left=1.1in,right=1.1in]{geometry} 
\usepackage{tcolorbox}
\usepackage[format=plain, labelfont={it,bf}, textfont=it, singlelinecheck=false]{caption} 
\usepackage[bottom]{footmisc}       
\usepackage{authblk}

\usepackage{color}
\usepackage{natbib}
\usepackage[hyphens]{url}
\usepackage{hyperref}
\hypersetup{
     colorlinks   = true,
     citecolor    = black,
     urlcolor     = blue,
     linkcolor    = black
}
\usepackage{doi}
\begin{document}

\title{Quantifying uncertainty about global and regional economic impacts of climate change}

\author[1,*]{Jenny Bjordal}
\author[1,2]{Trude Storelvmo}
\author[3]{Anthony A. Smith, Jr.}
\affil[1]{\footnotesize Department of Geosciences, University of Oslo, Oslo, Norway}
\affil[2]{\footnotesize Nord University, Bodø, Norway}
\affil[3]{\footnotesize Department of Economics, Yale University, New Haven, USA}
\affil[*]{\footnotesize e-mail: jenny.bjordal@geo.uio.no}
\renewcommand\Authands{ and }

\date{}

\maketitle

\begin{abstract}
    The economic impacts of climate change are highly uncertain. Two of the most important uncertainties are the sensitivity of the climate system and the so-called damage functions, which relate climate change to economic damages and benefits. Despite broad awareness of these uncertainties, it is unclear which of them is most important, both on the global as well as the regional level. Here we apply different damage functions to data from climate models with vastly different climate sensitivities, and find that uncertainty in both climate sensitivity and economic damage per degree of warming are of similar importance for the global economic impact. Increasing the climate sensitivity or the sensitivity of the damage function both increases the economic damages globally. Yet, at the country-level the effect varies depending on the initial temperature as well as how much the country warms. Our findings emphasise the importance of including these uncertainties in estimates of future economic impacts, as they both are vital for the resulting impacts and thus policy implications.

\end{abstract}

\section{Introduction}

Future projections of the economic impacts of climate change are highly uncertain. 
On the climate side, the most important uncertainty is how sensitive the climate system is to increasing amounts of greenhouse gases. 
On the economic side, the most important uncertainty is the relationship between climate change and economic impacts, often expressed in models by so-called damage functions. 
But which of these uncertainties is most important when assessing the economics impacts of future climate change?

A standard measure of how sensitive the climate is to greenhouse gas emissions is the equilibrium climate sensitivity (ECS) \citep{IPCC_WG1_2013}. 
This is how much the global mean surface air temperature will eventually increase when the amount of atmospheric CO$_2$ is doubled. For decades the estimated range for ECS has been 1.5-4.5$\degree$C \citep{IPCC_WG1_2013, charney1979carbon}. However, among the newest generation of climate models, there are several models with ECS above 5$\degree$C \citep[e.g.][]{zelinka2020causes}. 
Since economic impacts of climate change are currently mostly calculated from temperature \citep{vuuren2012comprehensive}, it is important to understand the implications of this uncertainty.

In economics, integrated assessment models (IAMs) for climate and economy are essential tools for simulating the economic impacts of climate change \citep[see e.g.][]{nordhaus1992optimal, ackerman2009limitations, weyant2017some}.
These models link climate and the economy by expressing changes in economic productivity (also called economic damages or benefits) as a function of the climate state, often represented by surface air temperature. So far, most damage functions have been applied to the global mean temperature \citep{nordhaus2018evolution, weitzman2012ghg} or to the temperature of large regions \citep{nordhaus1996regional, hope2011social, anthoff2014climate}, and therefore cannot tell us how individual countries or smaller regions will be impacted. 
The damage function is also an aspect of IAMs that has been criticised for being particularly uncertain \citep[see e.g.][]{ackerman2009limitations,weitzman2010what, howard2017few, diaz2017quantifying}, yet it is a crucial part of assessing the impacts of climate change.

\cite{hassler2018consequences} have previously found, using an IAM, that the importance of uncertainty in climate sensitivity and sensitivity in damage functions are of similar magnitude globally. Here, we aim to compare these two uncertainties using a different method, which also allows us to investigating their importance at the regional and country level.


We investigate the global and regional economic impacts of climate change by constructing regional damage functions that aggregate up to the global damages from already existing global damage functions.
Our approach builds on the method used by \cite{zheng2020climate}, by applying a damage function to climate model data, but our focus is on future projections dominated by greenhouse gas emissions, rather than aerosols. 
Additionally, we span the range of ECS and damage function uncertainties by using a low and a high climate sensitivity model, as well as a relatively insensitive and a relatively sensitive damage function. 

\section{Method}

\subsection{Climate model data and scenarios} 

We used data from a high-sensitivity model, the Community Earth System Model version 2 (CESM2, ECS=5.3$\degree$C)\citep{danabasoglu2020community}, and a low sensitivity model, the Norwegian Earth System Model (NorESM2, ECS=2.5$\degree$C)\citep{seland2020norwegian}.
Both participate in the 6th phase of the Coupled Model Intercomparison Project (CMIP6) \citep{eyring2015overview}, and together the two models span most of the ECS interval of 1.8–5.6 K found in CMIP6 \citep{zelinka2020causes}.

Specifically, the climate model data used were the surface air temperatures from the four future emission scenarios used in CMIP6 \citep{riahi2017shared, oneill2016scenario}, called shared socioeconomic pathways (SSPs). Since the scenarios go from 2015 to 2100, we also used the last years from runs with historical emission, because our base year here is year 2000. The scenarios are named using two numbers, where the first refers to the narrative (evolution of society and natural systems) \citep{oneill2014new}, and the second refer to the radiative forcing at the end of the century \citep{moss2010next}. Generally, lower (higher) numbers refer to lower (higher) emissions, and thus a lower (higher) temperature increase. A short summary of the scenarios are given in table \ref{table:scenarios}.

\begin{table}[!hbt]
\centering
    \begin{tabular}{| l | l | l | l |}
    \hline
        Scenario & Narrative & 2100 Forcing & Pathway of \\ 
        name & &  (W/m$^2$) & forcing \\ 
        \hline
        SSP1-2.6 & Sustainability & 2.6 & Peak and decline \\
        & (Low challenges to & & \\
        & mitigation and adaptation) & & \\
        SSP2-4.5 & Middle of the Road  & 4.5 & Stabilising \\
        & (Medium challenges to & & \\
        & mitigation and adaptation) & & \\
        SSP3-7.0 & Regional Rivalry & 7.0 & Rising \\
        & (High challenges to & & \\
        & mitigation and adaptation) & & \\
        SSP5-8.5 & Fossil-fueled Development & 8.5 & Rising \\
        & (High challenges to mitigation, & & \\
        & low challenges to adaptation) & & \\
        \hline
    \end{tabular}
\caption{Summary of the four future scenarios employed in this study. The narrative is what kind of world the scenario reflects, with indication of the mitigation and adaption challenges. The 2100 forcing gives the radiative imbalance at the end of the century under the scenario, and the pathway explains how the forcing evolves throughout the century.}
\label{table:scenarios}
\end{table}

\subsection{Damage functions and economic impacts} 

\subsubsection{Global damage functions}

Integrated assessment models typically incorporate global (or aggregate) economic damages from global warming as a reduction in global total factor productivity (TFP), a measure of the productivity of a set of factors of production (such as physical capital and labour) taken as a whole.  These damages can therefore be expressed as a fraction of global GDP, holding fixed productive inputs such as capital and labour, that varies with $\Delta T_t$, the change in the global surface air temperature in year $t$ from the pre-industrial temperature.  

That fraction, $\pi(\Delta T_t)$, is typically 
represented by an increasing convex function with the following functional form \citep[see e.g.][]{dietz2015endogenous}:
\begin{equation} \label{eq:global_damage}
    \pi(\Delta T_t) = \frac{\phi (\Delta T_t)^2}{1 + \phi (\Delta T_t)^2},
\end{equation}
where $\phi$ is a coefficient chosen to match estimates of aggregate damages from global warming. A change in the global temperature from $T_t$ to $T_{2000}$ would therefore lead to a percentage change in global TFP (and hence global GDP, holding inputs fixed) given by
\begin{equation} \label{eq:global_tfp_change}
    D(T_t) = \left( \frac{1+\phi(T_t - T_{1850})^2}{1+\phi(T_{2000} - T_{1850})^2} -1\right)100,
\end{equation}
where $T_{2000}$, $T_{1850}$ and $T_t$ is the global mean surface air temperature in years 2000, 1850 (pre-industrial), and year $t$, respectively.

\begin{figure}[htb!]
    \centering
    \includegraphics[width=1\textwidth]{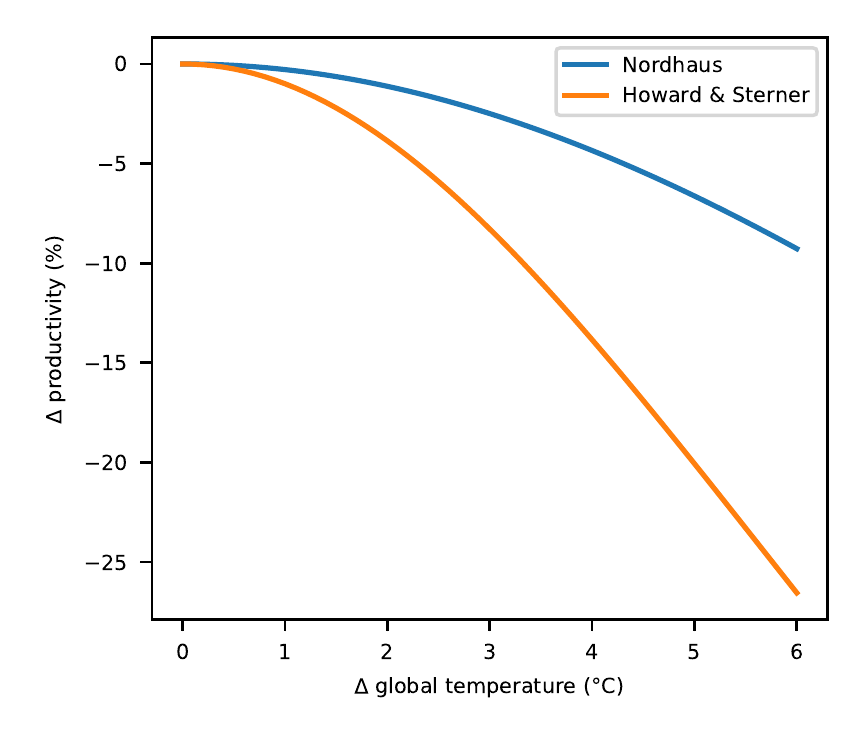}
    \caption{Global productivity change. The percentage change in productivity against temperature changes from pre-industrial for the Nordhaus (blue) and the Howard \& Sterner (orange) damage function.}
    \label{fig:compare_global_functions}
\end{figure}

The damage function developed by Nordhaus \citep{nordhaus1992optimal, nordhaus2018evolution} is one of the most commonly used damage functions and sets $\phi = 0.0028388$ (with unit 1/$\degree$C$^2$) (see blue line in figure \ref{fig:compare_global_functions}). Like many other damage functions, the Nordhaus damage function is only based on estimates of damages up to 3$\degree$C \citep{tol2011social, stern2013structure}, and is strictly speaking not valid for temperature increases beyond that. However, it is often extrapolated to higher temperature changes, and it is also criticised for being too optimistic, in that it may underestimate damages at large temperature increases \citep{stern2013structure, revesz2014improve}. We therefore use this as our low sensitivity damage function.

Another, more sensitive damage function, is the one estimated by \cite{howard2017few}, which sets $\phi = 0.0100380$ (see orange line in figure \ref{fig:compare_global_functions}). We use this as our high sensitivity damage function.
At 1$\degree$C warming, the difference in productivity between the two functions is less than 1\%, but as we approach 6$\degree$C warming the difference is more than 17\%, ranging from -9.3\% with the Nordhaus function to -26.5\% with the Howard \& Sterner function.

\subsubsection{Constructing regional damage functions}

In order to study how global warming affects regional economies, we constructed functions that capture how regional productivity varies with regional, rather than the global, surface air temperature. These functions are chosen so that the sum of regional variations across the globe matches estimates of global damages stemming from global warming. 
The full details on the calculations of the regional damage function is given in the appendix.

To operationalize the regional productivity function $H$, we used an inverse $U$-shaped function that depends on four parameters:
\begin{equation} \label{eq:regional_damage}
    H(T_{it}) =
    \begin{cases}
        (1 - b)e^{-\kappa^+(T_{it} - T^*)^2} + b & \text{if } T_{it} \geq T^* \\
        (1 - b)e^{-\kappa^-(T_{it} - T^*)^2} + b & \text{if } T_{it} < T^*,
    \end{cases}
\end{equation}
where $T_{it}$ is the temperature in region $i$ at time $t$, $T^*$ is the optimal temperature (given in $\degree$C) at which $H$ attains its maximum of 1, and $\kappa^+$ and $\kappa^-$ determine the steepness of the decline on either side of the optimal temperature.
The lower bound $b$ is set to 0.02. 
The remaining three parameters of the function H are chosen so that the aggregate damages from global warming implied by it match those delivered by the aggregate damage function D in equation \ref{eq:global_tfp_change} at three different global temperature changes ranging from 1 to 5$\degree$C, noting that the temperature in any particular region depends on the global temperature via a statistical downscaling model derived from runs of CESM2 and NorESM2 (see the appendix for details).
We used the downhill simplex equation to solve these three nonlinear equations in the three unknowns $T^{\ast}$, $\kappa^{+}$, and $\kappa^{-}$.
The resulting functions are shown in figure \ref{fig:regional_damage_functions} (and the parameters can be found in supplementary table \ref{table:df_parameters}). Note that the three parameters are unique for each climate model--damage function combination. And the resulting regional damages compared to the global damages are shown in supplementary figures \ref{fig:compare_temp} and \ref{fig:compare_scenarios}. As seen in the figures, the regional damage functions are not prefect fits for the global damage functions. But considering the large uncertainties for these functions, the resulting regional damage functions seem reasonable enough for our purpose.

\begin{figure}[htb!]
    \centering
    \includegraphics[width=1\textwidth]{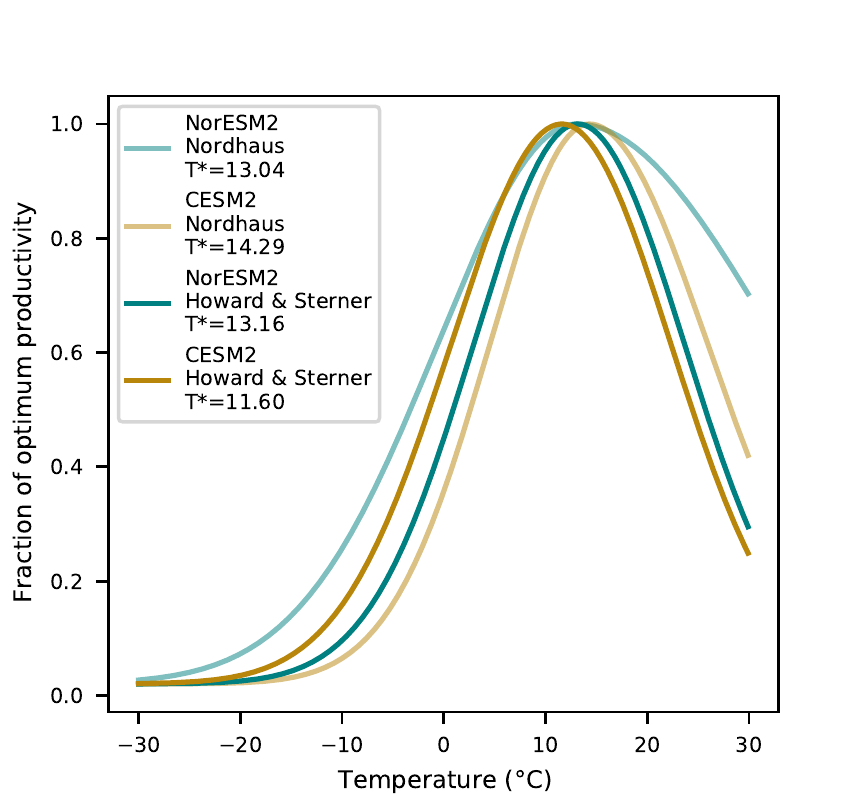}
    \caption{Regional damage functions for the four climate model--damage function combinations. Showing the fraction of optimum productivity against temperature, where fraction is one at the optimum temperature $T^*$.}
    \label{fig:regional_damage_functions}
\end{figure}

\subsection{Application of the damage function on the climate data}

After constructing the regional damage functions, we applied them to the climate model data. First, we calculated the annual spatial temperatures and re-gridded it to fit our 1$\degree \times $1$\degree$ population and GDP data grid. Then we calculated the fraction of optimum productivity for each year with the regional damage function. 

The 2000 temperature was calculated from a historical run, using the years 1996-2004. We used this to calculate the relative change in optimum productivity from 2000. Which was finally used to calculate the relative change in productivity.

\section{Results}

\subsection{Global economic impacts}

\begin{figure}[htb!]
    \centering
    \includegraphics[width=1\textwidth]{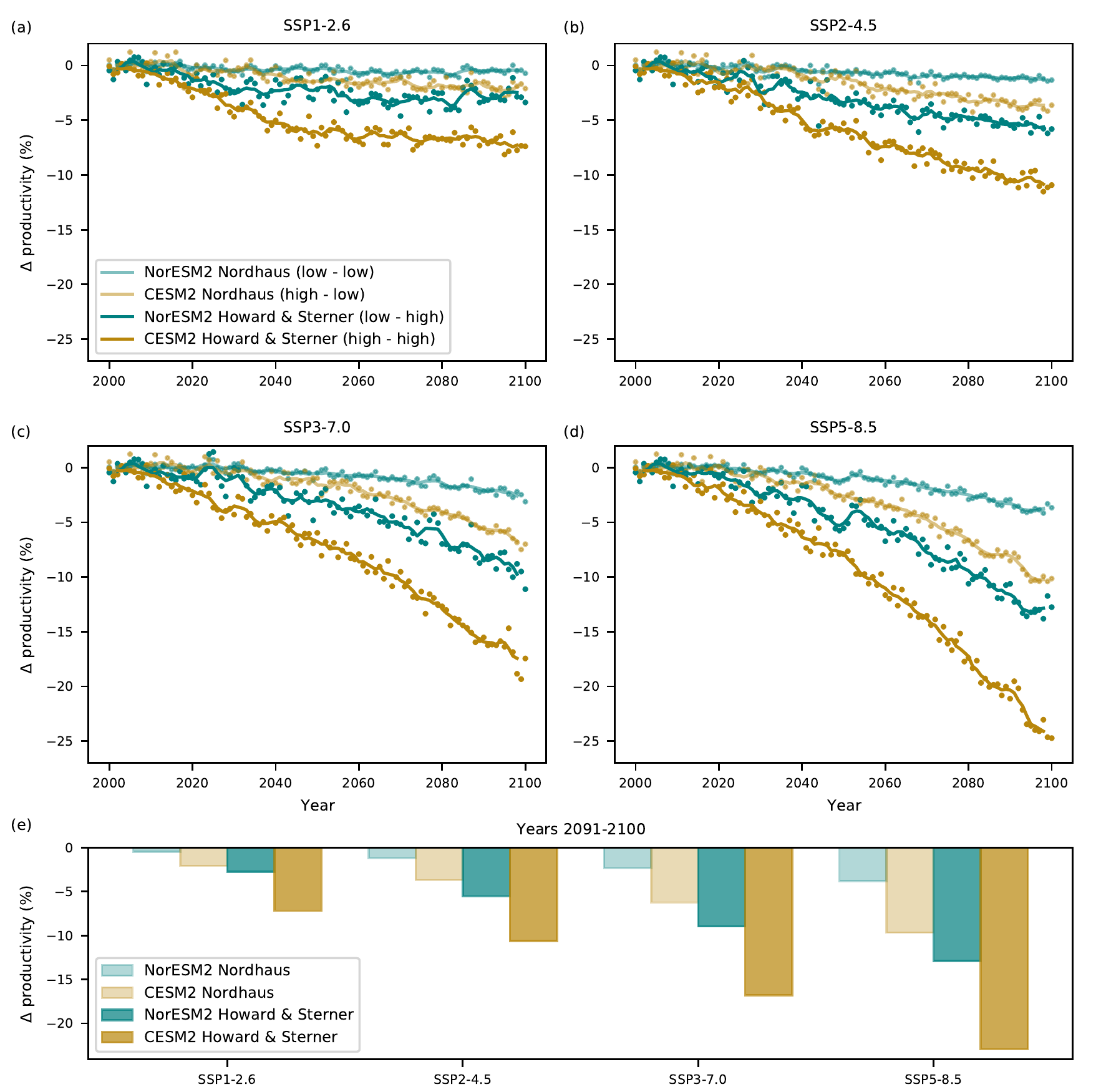}
    \caption{Percentage change in global productivity from year 2000 for the four SSPs. The line is the 5-year moving average, while the dots show each individual year's global productivity change from year 2000 for the four climate model--damage function combinations for SSP1-2.6 (a), SSP2-4.5 (b), SSP3-7.0 (c) and SSP5-8.5 (d). e, The change between the end of the century (years 2091-2100) and 2000 (1996-2004).}
    \label{fig:global_productivity}
\end{figure}

Our results show that global warming decreases the global economic productivity, but the size of these economic damages is highly dependent on the sensitivity of both the climate and the damage function (see fig. \ref{fig:global_productivity}). 
Both changing to a more sensitive climate model and a more sensitive damage function increases the damages. Also, the differences between models and damage functions grow as we reach higher temperatures, as seen from both the increasing distance between the lines with time in each SSP scenario, and the larger distance between lines in the higher emissions scenarios.
While the range in productivity due to the two damage functions shown here is slightly larger than that of the two climate models, the results indicate that the span of damages due to uncertainty in climate sensitivity and uncertainty associated with damage functions are of a similar magnitude globally, supporting previous work \citep{hassler2018consequences}.

\subsection{Regional economic impacts}

\begin{figure}[htb!]
    \centering
    \includegraphics[width=1\textwidth]{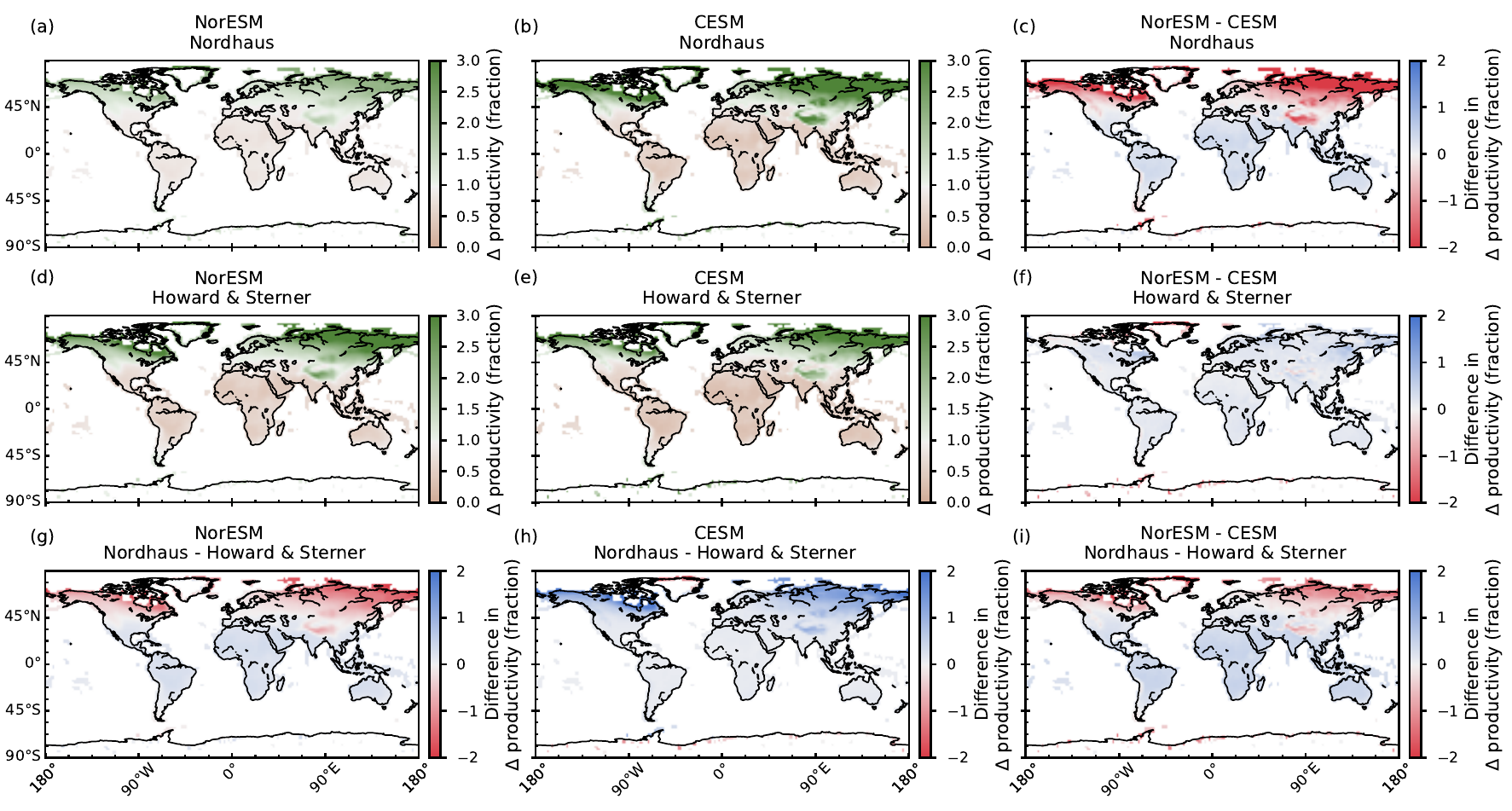}
    \caption{Spatial distribution of change in optimum productivity for SSP5-8.5. a,b,d,e Percentage change in optimum productivity at the end of the century (2091-2100) from year 2000 (1996-2004) for the four climate model--damage function combinations for SSP5-8.5. c,f,g,h,i The differences in productivity change between the combinations.}
    \label{fig:spatial_productivity}
\end{figure}

However, the global estimates of economic impacts of climate change hide a lot of heterogeneity at the regional level.
Regionally, our calculations predict decreasing productivity for most of the globe, except in the northern high latitudes and the Tibetan Plateau (fig. \ref{fig:spatial_productivity}). 
Figure \ref{fig:spatial_productivity} a,b,d,e shows how the fraction of the optimum productivity has changed from year 2000 to the end of the century in the SSP5-8.5 scenario for the four climate model--damage function combinations. The other scenarios have the same main patterns but of smaller magnitude (see supplementary figures \ref{fig:spatial_productivity_ssp126}-\ref{fig:spatial_productivity_ssp370}).  
While the area with increased productivity (green) might seem vast, much of the area has a small population and economy, and thus does not contribute much to the global productivity.

Also, changing between climate models and/or damage functions has different effects in different areas. 
If we start with the low--low combination (low climate sensitivity - low sensitivity damage function), going to a high sensitivity in either climate model (high--low, fig. \ref{fig:spatial_productivity} c) or damage function (low--high, fig. \ref{fig:spatial_productivity} g) increases the productivity in cold areas, and decreases the productivity in warm areas. But if we start with either the high sensitivity climate model (high--low) or damage function (low--high), and move to the high--high combination (fig. \ref{fig:spatial_productivity} f and h, respectively) the productivity decreases almost everywhere. 

This can be understood from the different temperature increases between models, and the different optimum temperatures and slopes of the regional damage functions (see figure \ref{fig:regional_damage_functions} and supplementary table \ref{table:df_parameters} for details).
When going from the low--low to the low--high combination we have the same warming and approximately the same optimum temperature. The larger increase in productivity in the cold regions and decrease in the warm regions is thus due to the steeper slopes of the damage function, making the regions move faster toward or away from the optimum temperature.
If we instead move from the low--low to the high--low combination we still have a similar optimum temperature and steeper slopes. Additionally, this effect is strengthened by the higher temperature increase.
In the other case, starting from the low--high or the high--low combination and going to the high--high combination, we have a different situation. Now the slopes are not changing much, but the optimum temperature is lower for the high--high combination. This means that fewer regions have the potential to increase their productivity, and more regions will cross over the optimum temperature and start decreasing their productivity. Additionally, the baseline temperature difference between the models plays a part when changing model.

\subsection{Large variation between countries}

\begin{figure}[htb!]
    \centering
    \includegraphics[width=1\textwidth]{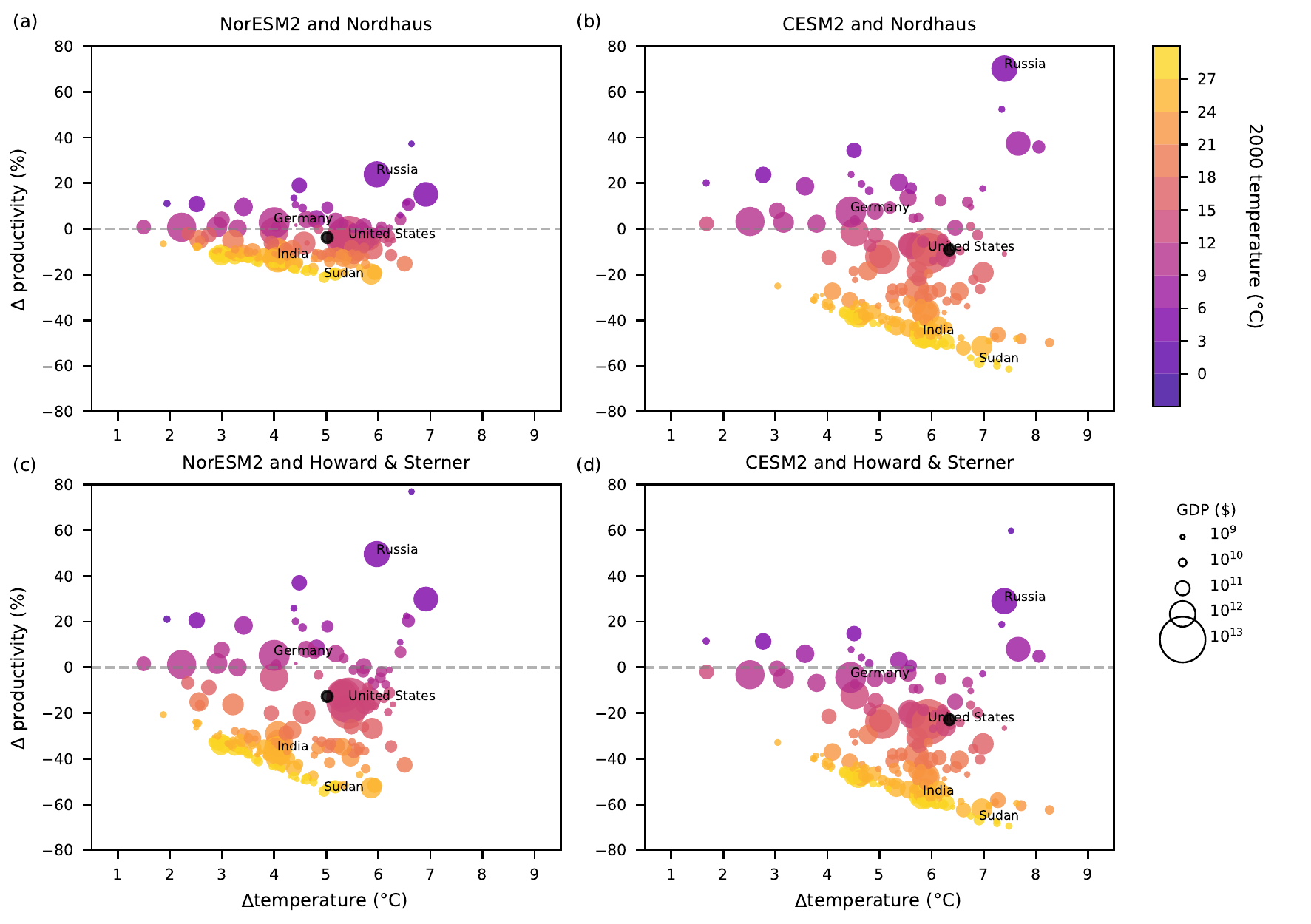}
    \caption{Country-level temperature and productivity change. Showing the four climate model--damage function combinations' productivity change at end of the century (2091-2100) from year 2000 (1996-2004) against population-weighted temperature change for SSP5-8.5. Each country's dot is coloured based on the year 2000 population-weighted temperature, and the size indicates the GDP in year 2000. The black dot is the global average (and does not indicate temperature or GDP).}
    \label{fig:country_productivity}
\end{figure}

On the country level, as indicated by figure \ref{fig:spatial_productivity}, it is the countries with a cool climate that benefit, while the warm countries suffer damages. 
Figure \ref{fig:country_productivity} shows how the temperature and productivity change from 2000 to the end of the century for each country in each climate model--damage function combination for SSP5-8.5 (for the other SSPs see supplementary figures \ref{fig:country_productivity_ssp126}-\ref{fig:country_productivity_ssp370}). Like the global average, the country-level average takes into account each country's economic activity, measured in gross domestic product (GDP) (indicated by size of circle), and population in year 2000. The colour of the circles show the population-weighted temperature in year 2000. Again, the same three factors that explain figure \ref{fig:spatial_productivity} are important for each country's change in productivity, as well as the countries' distribution of population and economic activity.
We see that most countries will experience economic impacts that are very different from the global average (black dot).

\begin{figure}[htb!]
    \centering
    \includegraphics[width=0.65\textwidth]{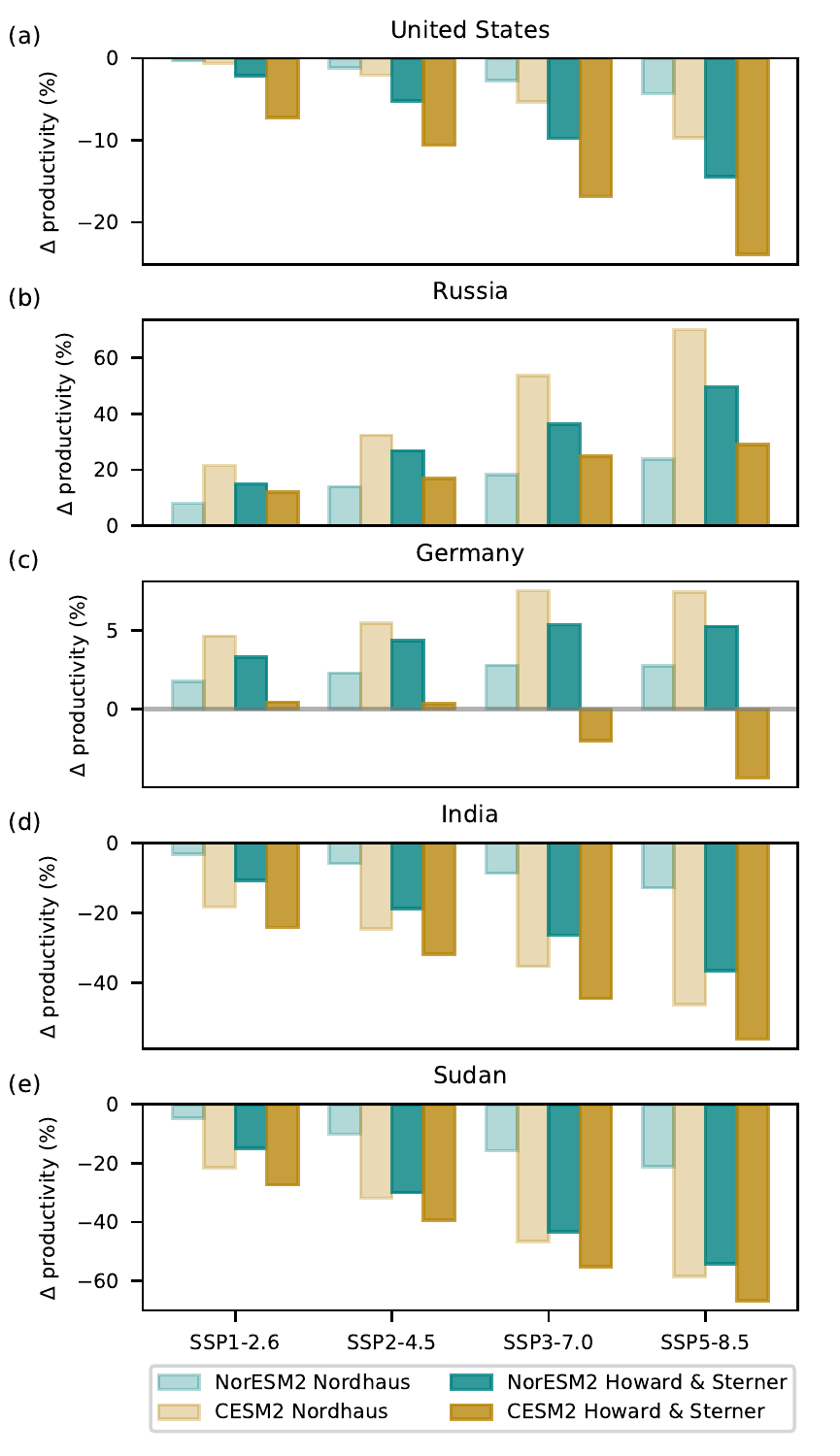}
    \caption{Country-level productivity change. Showing the percentage change in productivity at the end of the century (2091-2100) from year 2000 (1996-2004) for the four climate model--damage function combinations and the four future SSP scenarios for five selected countries.}
    \label{fig:selected_countries}
\end{figure}

Some examples of how countries may be impacted in each scenario and climate model--damage function combination are shown in fig. \ref{fig:selected_countries}. The United States, the biggest country-level economy, nicely follows what we saw for the global productivity (fig. \ref{fig:global_productivity} e), while other countries show very different responses. Russia is one of the countries that benefit from climate change independent of scenario and combination, while India and Sudan clearly see economic damages under all circumstances. However, these countries do not follow the same increase in effect with increasing sensitivity as seen globally and for the United States. 
On the regional level it is not clear that increasing the climate sensitivity and/or sensitivity of the damage function increases the economic impacts. 

Another interesting group of countries is the one with relatively small impacts that lie close to the zero line in figure \ref{fig:country_productivity}. Germany (fig. \ref{fig:selected_countries} c) is a nice example of these, and we see that whether the country will see benefits or damages due to climate change depends on both the scenario and the climate model--damage function combination. This is because Germany has a temperature lower than, yet close to, the optimal temperature (10.0$\degree$C in NorESM2, 10.4$\degree$C in CESM2) in year 2000. As the climate warms, as decided by both the scenario and the climate model, Germany might cross the optimum temperature. 
When crossing the optimum temperature, the productivity starts to decrease, and if moving too far away from the optimum temperature the end productivity could be lower than at the start point in year 2000, depending on the slopes of the damage function.

\section{Discussion}

Our results show how the large uncertainties in climate sensitivity and damage functions result in large uncertainties in economic impacts. This points out how important it is to include uncertainty estimates when calculating economic impacts of climate change. Particularly, this is important when using estimates of economic impacts for policy purposes, like for example the US government does  \citep{auffhammer2018quantifying}.
The study also points out how different the economic impacts of climate change can be between regions. Especially, we see how the regional uncertainty in economic impacts can be very different from the global.

In a full economic model, the economic impacts will feed back on the emissions. With economic damages, consumption will decrease, and consequently the emissions (and thus the warming) will be lower (and visa versa for economic benefits). In our work, future emissions are already set by the SSP scenarios, and this climate-economy feedback is therefore not included. However, the study still gives important insights into the relative importance of the climate sensitivity and sensitivity of the damage function. Both of these are important for the size of the climate-economy feedback, and this could be a step towards assessing the feedback's uncertainty.

Our calculations are based on the assumption that the population growth is constant in time and space, and the actual future population growth and migration could therefore result in a different future global productivity. If, for example, people living in less productive areas move northward into more productive areas, the global productivity would decrease less with warming. Such migration to more productive areas are indeed expected, and even simulated in some models \citep{conte2020local}.

An important remaining question is how good of a proxy temperature is for climate change. In the damage functions employed here, it is assumed that the global temperature represents global climate change well. This is probably a reasonable assumption, but it might not hold as well when we construct regional damage functions that look at regional temperature. Changes in other climate variables, like rainfall, sea level rise or extreme events, do not necessarily have the same distribution patterns as temperature change \citep{IPCC_WG1_2013}.  Sea-level rise, for example, should only affect regions with a coast line, but when temperature is the only input to the damage function, the inherent damages from sea-level rise in the global damage function will be distributed depending on temperature in the regional function. Thus the impacts of climate change that do not follow the temperature distribution could be wrongly distributed.

In our study, we have used damage functions that calculate the impact of climate change on the economic output rather than on the economic growth rate. 
Yet, this is an area of some debate \citep[see e.g.][]{moore2015temperature, burke2015global}. Since even a small change in the growth rate could lead to large economic impacts in the long run, such damage functions could be even more sensitive than the high sensitivity damage function employed here.

\section{Conclusion}

We have found that the uncertainties in climate sensitivity and damage functions are of similar magnitude globally, but differ a lot regionally.
What is clear from this study is that progress toward reliable assessments of economic damages due to climate change will require both more constrained estimates of the climate sensitivity and improved damage functions. Especially, this is important when economic models are used for policy purposes \citep[see e.g.][]{burke2016opportunities, hassler2018consequences}.
We see that the global productivity tells us little about what will happen regionally or in a given country. The regional productivity could also be more impacted by changes that are not as well represented by the annual mean surface air temperature, such as rainfall or extreme events. A focus on constructing regional damage functions will therefore be important.

\clearpage

\subsubsection*{Acknowledgements}
This work was supported by the Norwegian Research Council through grant 281071. 
We acknowledge the World Climate Research Programme, which, through its Working Group on Coupled Modelling, coordinated and promoted CMIP6. We thank the climate modelling groups for producing and making available their model output, the Earth System Grid Federation (ESGF) for archiving the data and providing access, and the multiple funding agencies who support CMIP6 and ESGF.


\subsubsection*{Competing Interests statement}
The authors declare no competing interests.

\subsubsection*{Data availability}
All NorESM2 and CESM2 simulation output is available through the Earth System Grid Federation's (ESGF) CMIP6 search interface (\url{https://esgf-node.llnl.gov/search/cmip6/}) under ‘source ID’ \textit{NorESM2-LM} and \textit{CESM2}, respectively, and ‘experiment ID’ \textit{historical}, \textit{ssp126}, \textit{ssp245}, \textit{ssp370} and \textit{ssp585}. \\
Population and GDP data from the Geographically based Economic data (G-Econ) site, version 2.11 \citep{nordhaus2006gecon}, is no longer available on the web site, but is available from the authors on request. 

\bibliography{referanser}

\clearpage
\setcounter{page}{1}

\section*{Supplementary Material}

\renewcommand{\figurename}{Supplementary Figure}
\setcounter{figure}{0}
\renewcommand{\tablename}{Supplementary Table}
\setcounter{table}{0}

\appendix
\section*{Appendix}
To look at regional differences in productivity stemming from global warming, 
we needed a regional function mapping regional surface air temperatures into regional productivity.  Following \cite{krusell2020climate}, we constructed such a function so that  regional variations in productivity, when summed across all regions, match the global changes in productivity delivered by the aggregate damage function (equation \ref{eq:global_tfp_change}).

To do this, we used a simple economic model of regional GDP:
\begin{equation} \label{eq:regional_gdp}
    Y_{it} = K_{it}^{\alpha} L_{it}^{1-\alpha},
\end{equation}
where, in region (or grid-cell) $i$ in year $t$, $Y_{it}$ is GDP,
$K_{it}$ is the physical capital stock, and $L_{it}$ is the effective supply of labour, measured in so-called ``efficiency units'' which capture how productive workers are.  The coefficients $\alpha$ and $1-\alpha$ are the shares of income (GDP) going to capital and labour, respectively.  We assume further that the effective supply of labour evolves according to:
\begin{equation} \label{eq:effective}
L_{it} = A_{it} H(T_{it}) N_{it},
\end{equation}
where, in region $i$ in year $t$, $N_{it}$ is the population and $A_{it} H(T_{it})$ is the number of efficiency units per person.  Population is assumed to grow at rate $n$.  The first component of the number of efficiency units, $A_{it}$, does not depend on regional temperature and is assumed to grow at rate $g$. The second component of the number of efficiency units, $H(T_{it})$, does depend on regional temperature: it captures on how changes in regional temperature, $T_{it}$, affect the regional productivity of labour.  The function $H$ governing this component of regional productivity has an inverse $U$-shape and is bounded between 0 and 1.

Finally, we assume that at any point in time capital is freely mobile across regions so that the marginal product of capital (i.e., the extra amount of GDP generated by an incremental amount of additional capital) is equated across regions.  Consequently, regional capital-to-labour ratios, $K_{it}/L_{it}$, are also equalised, implying in turn that global GDP in year $t$, $Y_t \equiv \sum_{i=1}^M Y_{it}$, where $M$ is the number of regions, has a simple expression:
\begin{equation} \label{eq:globalgdp}
Y_t = K_t^{\alpha} \left(\sum_{i=1}^M L_{it}\right)^{1-\alpha},
\end{equation}
where $K_t \equiv \sum_{i=1}^M K_{it}$ is the global capital stock.\footnote{\cite{krusell2020climate} show that even when capital markets are closed completely optimal accumulation of capital in each region leads the marginal product of capital to be approximately equalised across regions.  Thus the assumption of free capital mobility underlying the formulas in this paper appears to be an innocuous one.}
Substituting equation (\ref{eq:effective}) into equation (\ref{eq:globalgdp}) yields an expression for  global GDP in year $t$ depends explicitly on the set of regional temperatures:
\begin{equation} \label{eq:globalgdp2}
Y_t = K_t^{\alpha} \left(\sum_{i=1}^M ((1+g)(1+n))^{t-2000} A_{i,2000} \, N_{i,2000} \, H(T_{it})\right)^{1-\alpha}.
\end{equation}

To derive an analogous expression for global GDP that depends only on the global temperature, we used a statistical downscaling model that relates regional temperature to global temperature:
\begin{equation} \label{eq:regional_temperature}
    T_{it} = T_{i,2000} + \gamma_i(T_t - T_{2000}).
\end{equation}
To obtain the region-specific responsiveness coefficients $\gamma_i$, we proceeded in three steps.
First, we calculated the global temperature change from pre-industrial to year 2000, which is our reference year. 
Second, we calculated the temperatures in year 2000 for each grid cell ($T_{i,2000}$). Here we used surface air temperature from three historical runs by the climate model, and calculated the average of these three runs using the nine years around 2000 (1996-2004).
Finally, we calculated how each grid cell's temperature changes relative to the global temperature, the responsiveness coefficients ($\gamma_i$). These we calculated by using five different model runs: historical, SSP1-2.6, SSP2-4.5, SSP3-7.0 and SSP5-8.5. For each run, the difference between the first and the last five years was calculated, and each grid cell's change was divided by the global temperature change. The average of these five coefficients was then used as the final set of responsiveness coefficients for each of the two models (e.i. NorESM2 and CESM2). 

Armed with the statistical downscaling model in equation (\ref{eq:regional_temperature}), equation (\ref{eq:globalgdp2}) can now be rewritten:
\begin{equation} \label{eq:globalgdp3}
Y_t = S_t \, G(T_t) \, K_t^{\alpha} 
\end{equation}
where $S_t \equiv \left[(1+g)(1+n))^{t-2000}\right]^{1-\alpha}$, 
$G(T_t) \equiv
\left(\sum_{i=1}^M a_i \, N_{i,2000} 
{H(T_{i,2000} + \gamma_i(T_t - T_{2000})) \over H(T_{i,2000})}\right)^{1-\alpha}$, 
and $a_{i,2000} \equiv A_{i,2000} H(T_{i,2000})$ is efficiency units per person in region $i$ in 2000.
Given a regional productivity function $H$ (in our case equation \ref{eq:regional_damage}), a change in the global temperature from $T_t$ to $T_{2000}$ would lead to a percentage change in global GDP (again holding inputs fixed) equal to:
\begin{equation} \label{eq:global_gdp_change}
    d(T_t) \equiv \left({G(T_{2000}) \over G(T_t)}-1\right) 100.
\end{equation}

The goal now is to choose $H$ (equation \ref{eq:regional_damage}) so that the two ``damage functions'', $D(T_t)$ and $d(T_t)$, the first taken from existing estimates of global damages from climate change and the second derived from the simple economic model of regional damages from climate change outlined here, agree for different values of the global temperature $T_t$.

Regional population and GDP data for the year 2000 are taken from the G-Econ database, version 2.11 \citep{nordhaus2006gecon}.  The $a_{i,2000}$ are chosen by solving the following two equations for $K_{i,2000}$ and $a_{i,2000}$ in each region:
\begin{eqnarray*}
K_{i,2000}^{\alpha} \, (a_{i,2000} \, N_{i,2000})^{1-\alpha} & = & Y_{i,2000} \\
\alpha K_{i,2000}^{\alpha-1} \, (a_{i,2000} \, N_{i,2000})^{1-\alpha} & = & r.
\end{eqnarray*}
The first of these equations ensures that regional GDP in year 2000 equals its value in the data and the second of these equations imposes that the marginal product of capital in each region is equated to a common rate of return $r$ (net of depreciation), here set to 2.53\%.
Capital's share of income, $\alpha$, is set to 0.36.

\begin{table}[!hbt]
\centering
    \begin{tabular}{| l | l | l l l |}
    \hline
        Damage function & Model & T$^*$ & $\kappa^-$ & $\kappa^+$ \\ 
        \hline
        Nordhaus & NorESM2 & 13.0 & 0.00267 & 0.00127 \\
        Nordhaus & CESM2 & 14.3 & 0.00522 & 0.00367 \\
        Howard \& Sterner & NorESM2 & 13.2 & 0.00476 & 0.00453 \\
        Howard \& Sterner & CESM2 & 11.6 & 0.00420 & 0.00434 \\
        \hline
    \end{tabular}
\caption{The parameter values found for the four climate model--damage function combinations.}
\label{table:df_parameters}
\end{table}

\begin{figure}[htb!]
    \centering
    \includegraphics[width=1\textwidth]{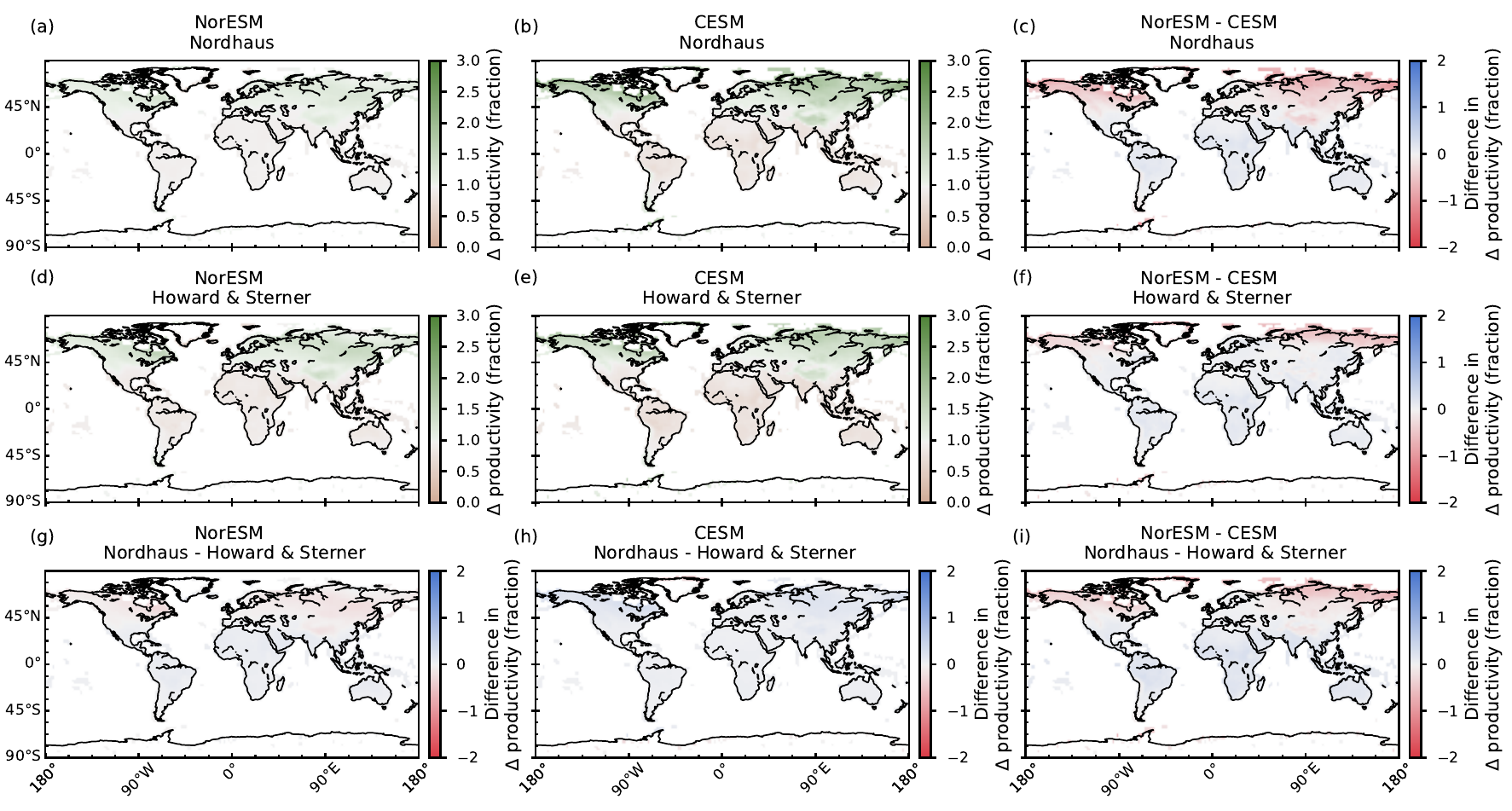}
    \caption{Country-level temperature and productivity change for SSP1-2.6. Showing the four climate model--damage function combinations' productivity change at end of the century (2091-2100) from year 2000 (1996-2004) against population-weighted temperature change for SSP1-2.6. Each country's dot is coloured based on the year 2000 population-weighted temperature, and the size indicates the GDP in year 2000. The black dot is the global average (and does not indicate temperature or GDP).}
    \label{fig:spatial_productivity_ssp126}
\end{figure}

\begin{figure}[htb!]
    \centering
    \includegraphics[width=1\textwidth]{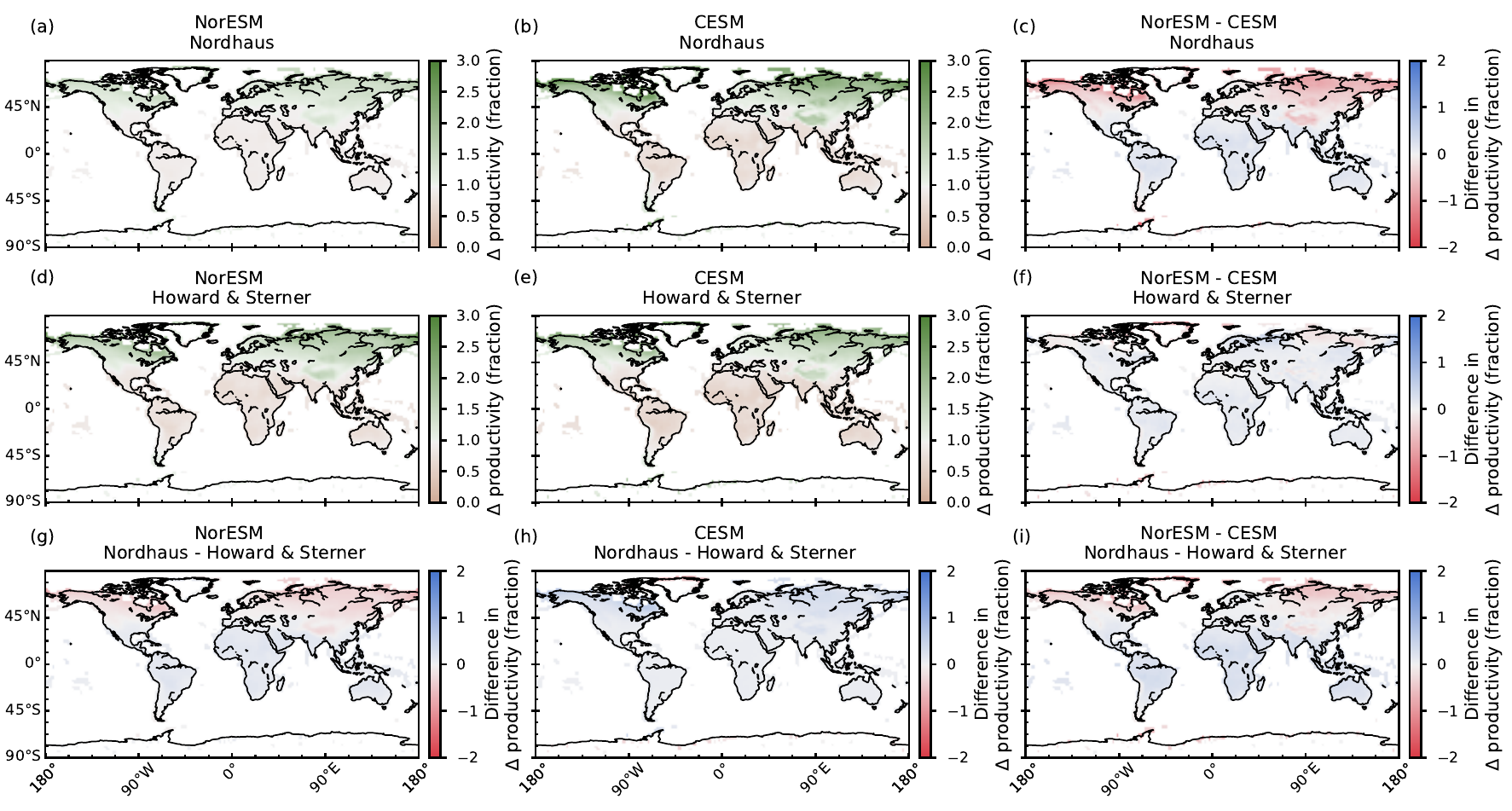}
    \caption{Country-level temperature and productivity change for SSP2-4.5. Same as figure \ref{fig:spatial_productivity_ssp126}, but for SSP2-4.5.}
    \label{fig:spatial_productivity_ssp245}
\end{figure}

\begin{figure}[htb!]
    \centering
    \includegraphics[width=1\textwidth]{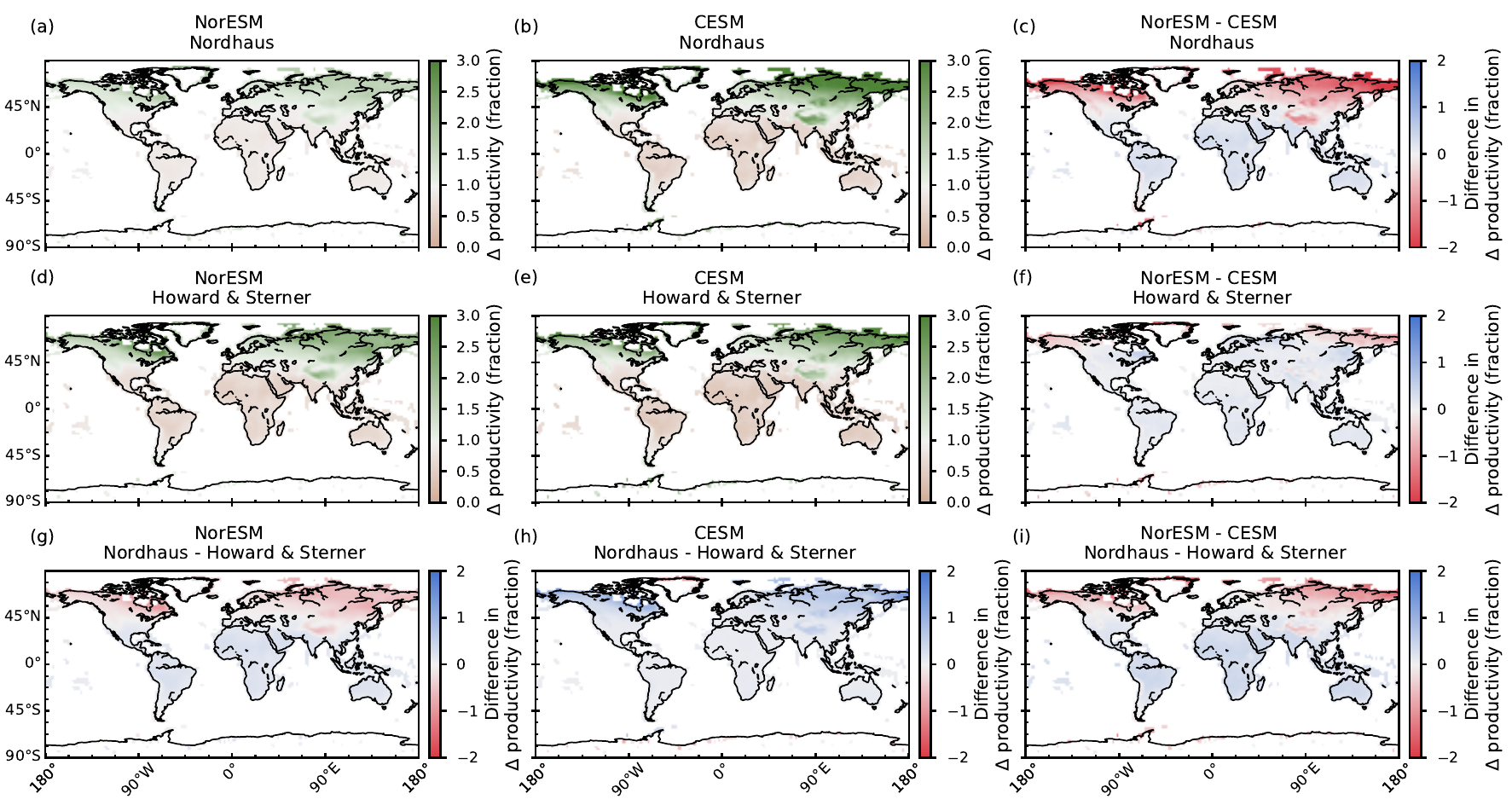}
    \caption{Country-level temperature and productivity change for SSP3-7.0. Same as figure \ref{fig:spatial_productivity_ssp126}, but for SSP3-7.0.}
    \label{fig:spatial_productivity_ssp370}
\end{figure}

\begin{figure}[htb!]
    \centering
    \includegraphics[width=1\textwidth]{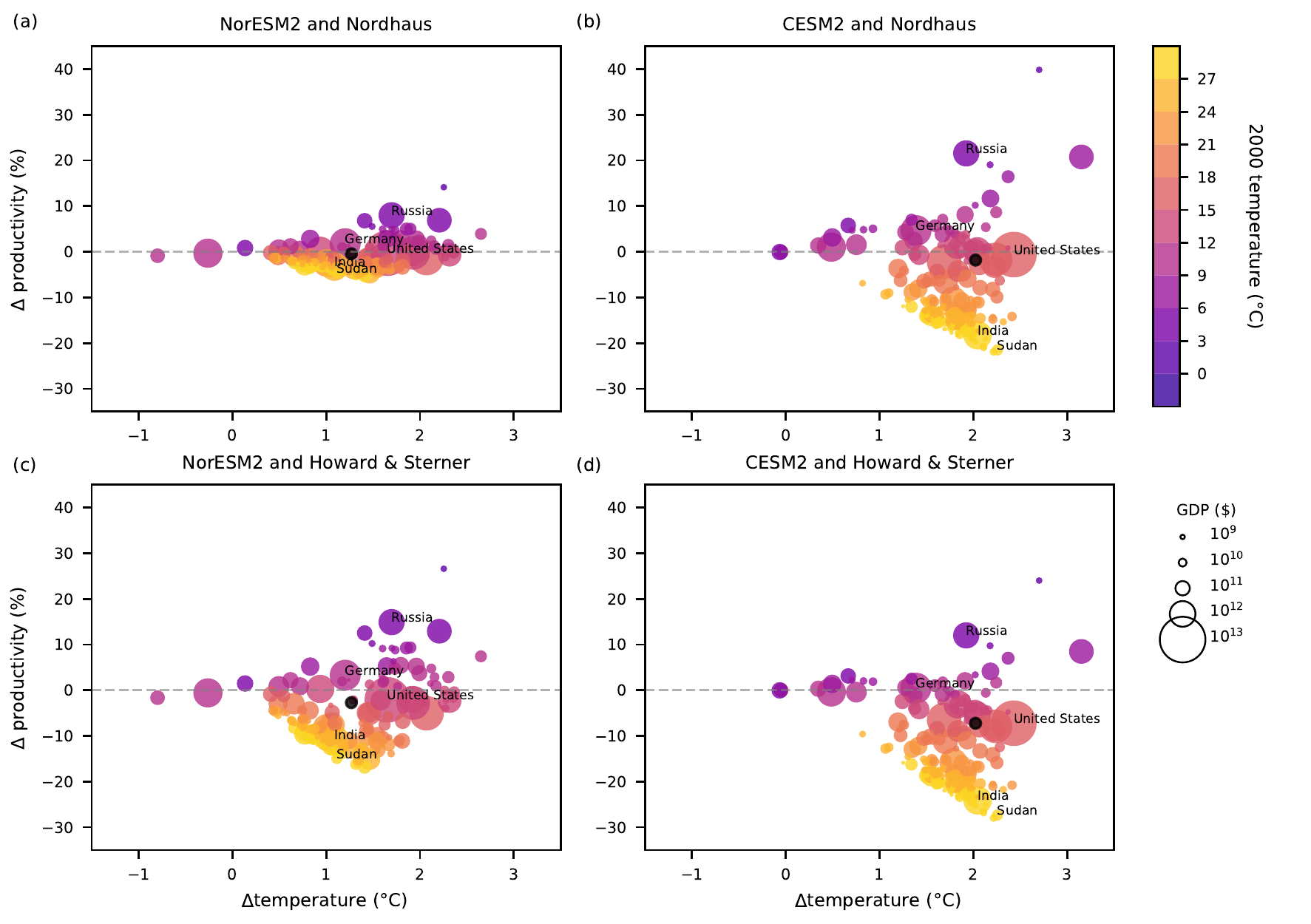}
    \caption{Country-level temperature and productivity change for SSP1-2.6. Showing the four climate model--damage function combinations' productivity change at end of the century (2091-2100) from year 2000 (1996-2004) against population-weighted temperature change for SSP1-2.6. Each country's dot is coloured based on the year 2000 population-weighted temperature, and the size indicates the GDP in year 2000. The black dot is the global average (and does not indicate temperature or GDP).}
    \label{fig:country_productivity_ssp126}
\end{figure}

\begin{figure}[htb!]
    \centering
    \includegraphics[width=1\textwidth]{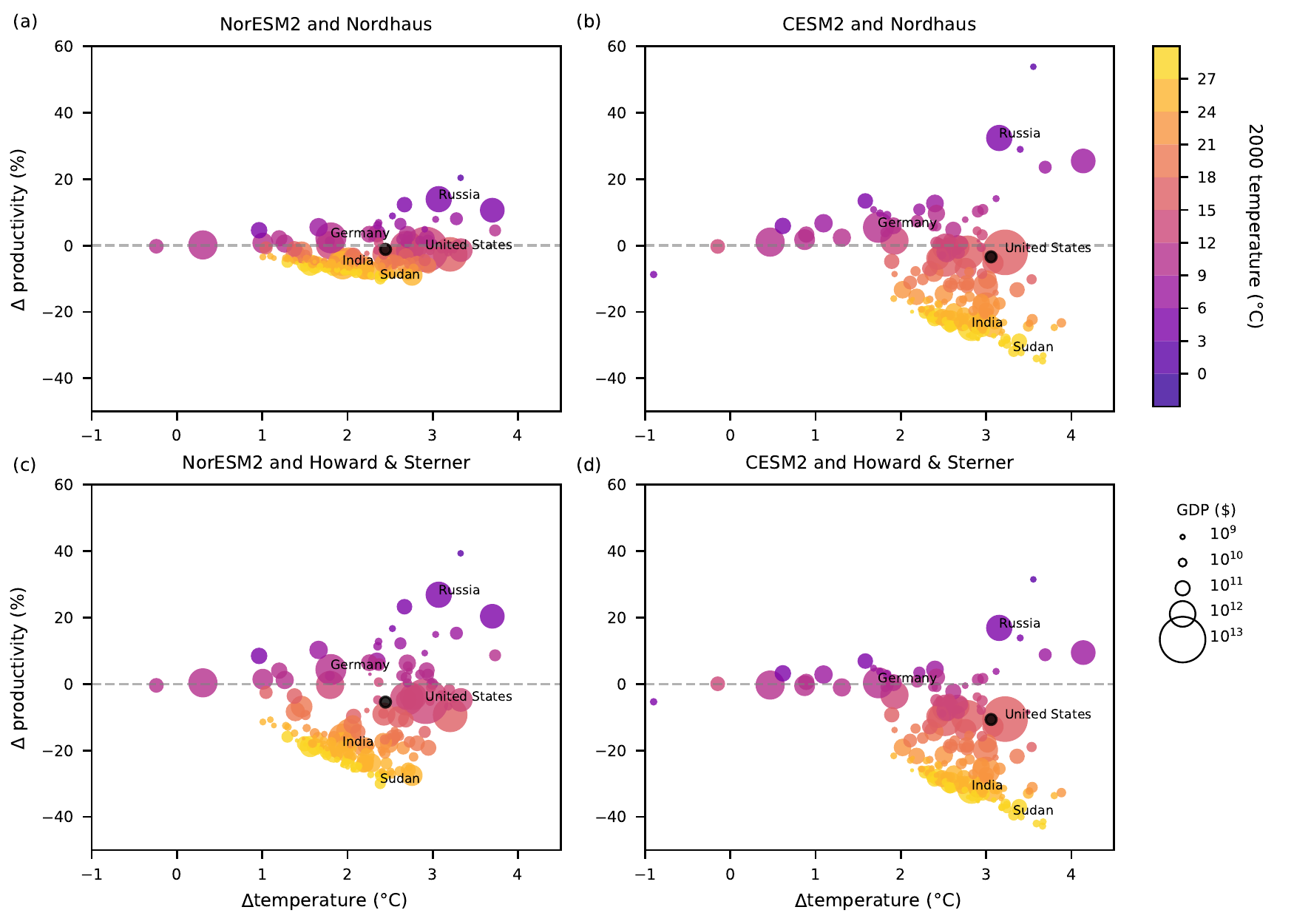}
    \caption{Country-level temperature and productivity change for SSP2-4.5. Same as figure \ref{fig:country_productivity_ssp126}, but for SSP2-4.5.}
    \label{fig:country_productivity_ssp245}
\end{figure}

\begin{figure}[htb!]
    \centering
    \includegraphics[width=1\textwidth]{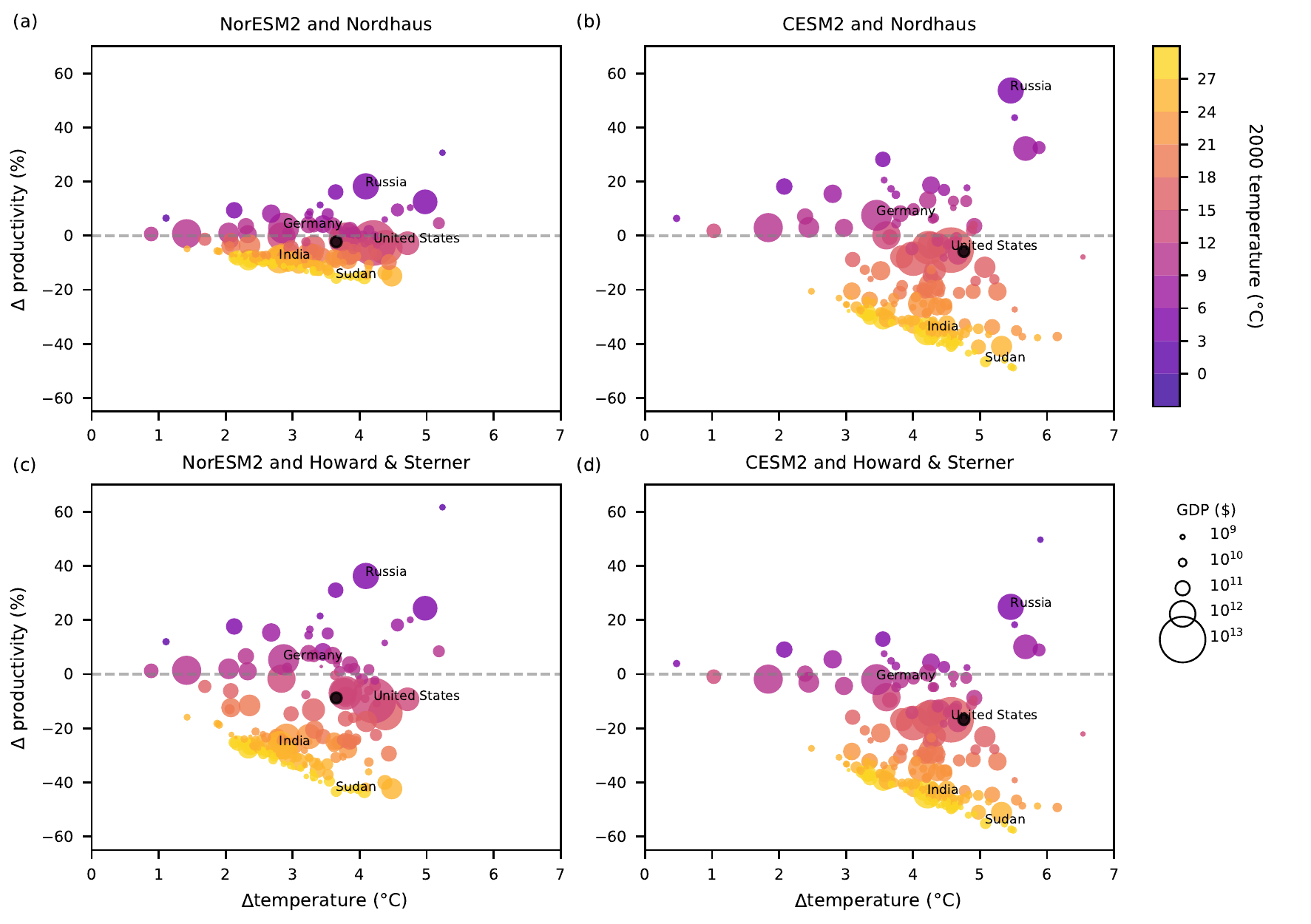}
    \caption{Country-level temperature and productivity change for SSP3-7.0. Same as figure \ref{fig:country_productivity_ssp126}, but for SSP3-7.0.}
    \label{fig:country_productivity_ssp370}
\end{figure}

\begin{figure}[htb!]
    \centering
    \includegraphics[width=1\textwidth]{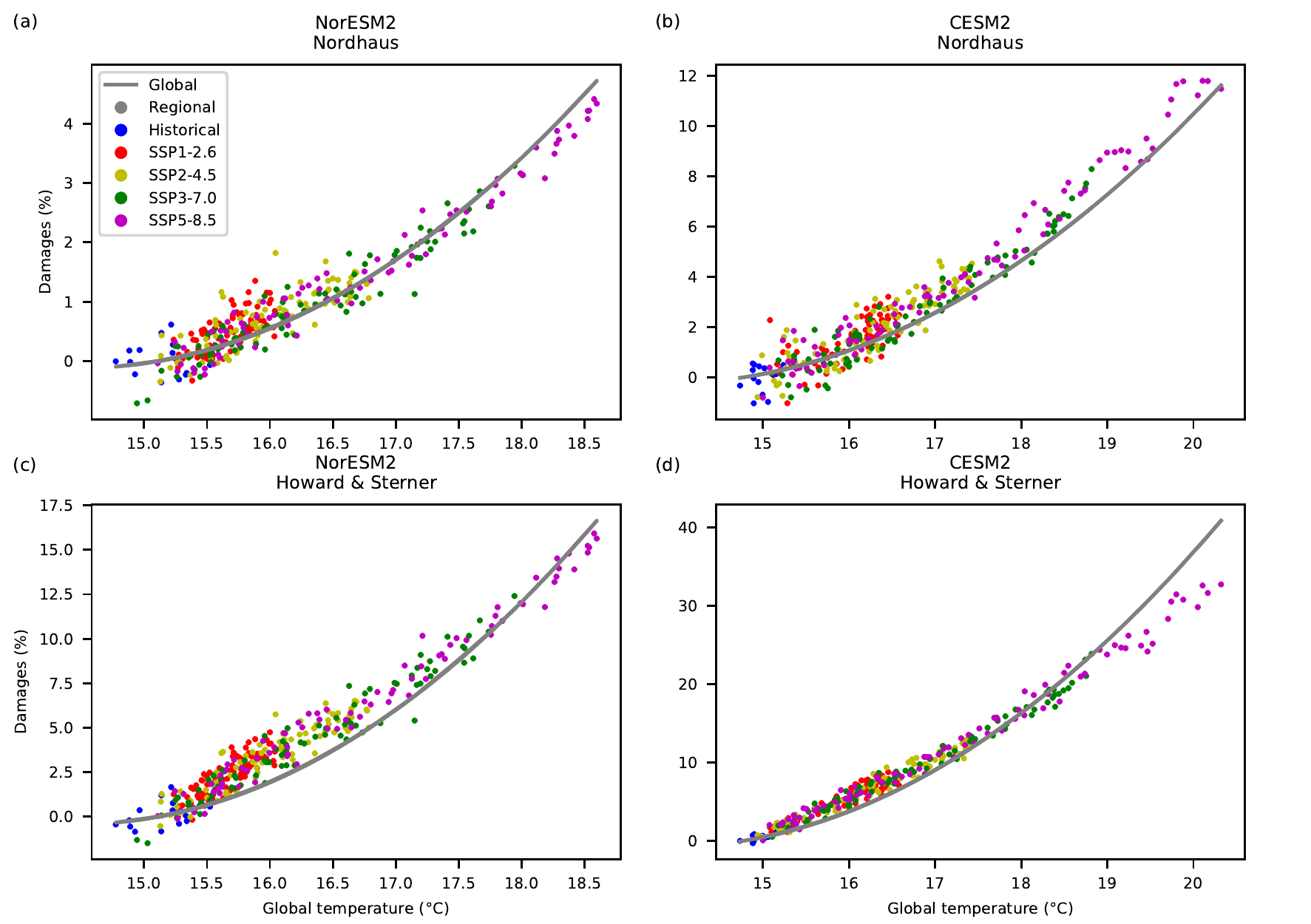}
    \caption{Global and regional damages plotted against temperature. Each year from historical run and the SSP scenarios are plotted for both the global (line) and the regional (dots), for the four climate model damage function combinations.}
    \label{fig:compare_temp}
\end{figure}

\begin{figure}[htb!]
    \centering
    \includegraphics[width=1\textwidth]{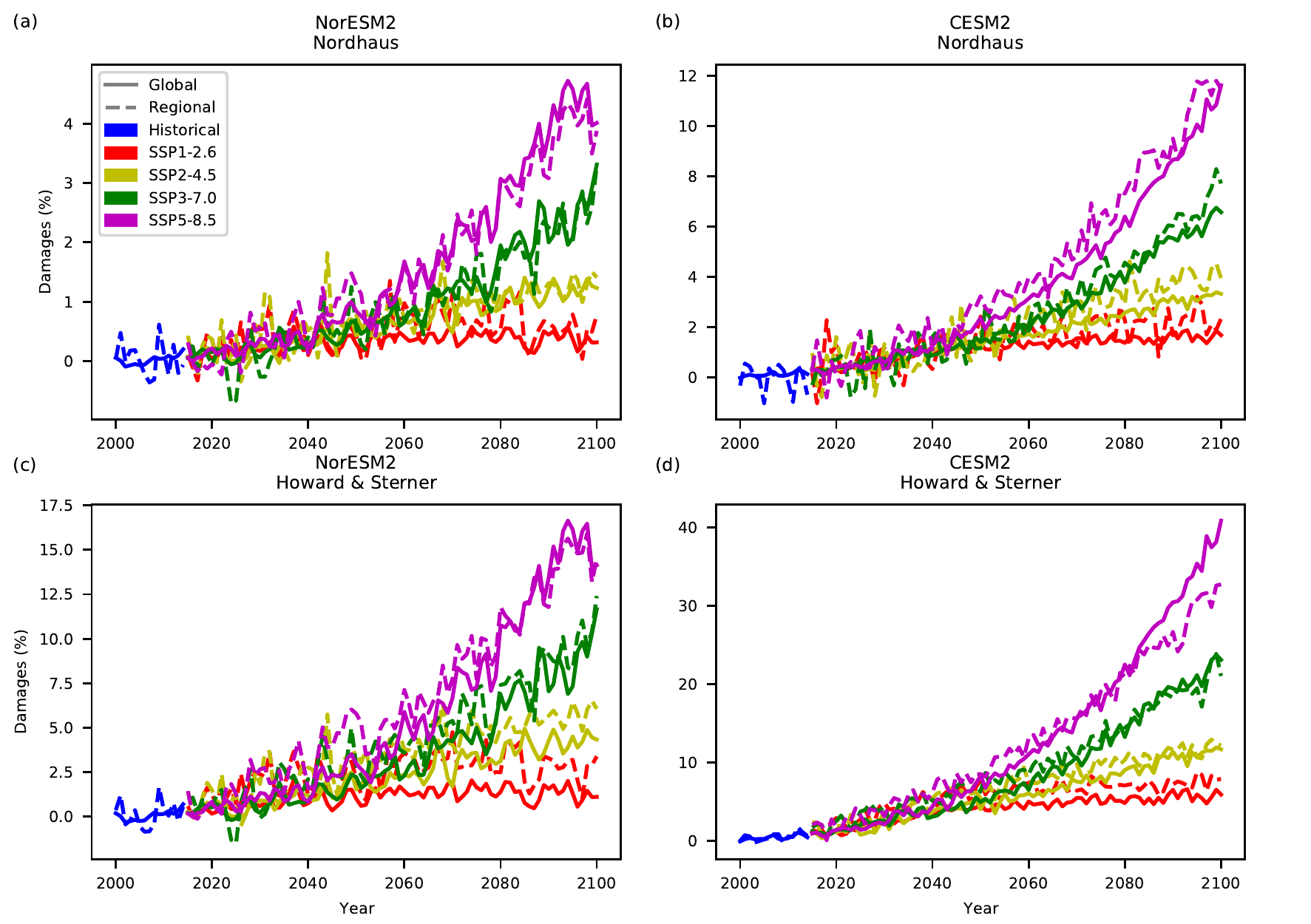}
    \caption{Global and regional damages plotted against time.} Each year from historical run and the SSP scenarios are plotted for both the global (line) and the regional (dashed), for the four climate model damage function combinations.
    \label{fig:compare_scenarios}
\end{figure}

\end{document}